\documentclass[prb,showpacs,twocolumn,preprintnumbers,amsmath,groupedaddress,amssymb,superscriptaddress,floatfix]{revtex4}
\usepackage[dvips]{graphicx}
\usepackage[latin1]{inputenc}
\usepackage{xcolor,bm}
\usepackage{nicefrac}

\begin{document}
\draft

\hyphenation{a-long}

\title{Mutual independence of critical temperature and superfluid density\\ under pressure in optimally electron-doped superconducting LaFeAsO$_{1-x}$F$_{x}$}

\author{G.~Prando}\email[E-mail: ]{g.prando@ifw-dresden.de}\affiliation{Leibniz-Institut f\"ur Festk\"orper- und Werkstoffforschung (IFW) Dresden, D-01171 Dresden, Germany}
\author{Th.~Hartmann}\affiliation{Institut f\"ur Theoretische Physik III, Ruhr-Universit\"at Bochum, D-44801 Bochum, Germany}
\author{W.~Schottenhamel}\affiliation{Leibniz-Institut f\"ur Festk\"orper- und Werkstoffforschung (IFW) Dresden, D-01171 Dresden, Germany}
\author{Z.~Guguchia}\affiliation{Laboratory for Muon Spin Spectroscopy, Paul Scherrer Institut, CH-5232 Villigen PSI, Switzerland}
\author{S.~Sanna}\affiliation{Dipartimento di Fisica and Unit\`a CNISM di Pavia, Universit\`a di Pavia, I-27100 Pavia, Italy}
\author{F.~Ahn}\affiliation{Institut f\"ur Theoretische Physik III, Ruhr-Universit\"at Bochum, D-44801 Bochum, Germany}
\author{I.~Nekrasov}\affiliation{Institute for Electrophysics, Russian Academy of Sciences, Ural Branch, Amundsen str. 106, Ekaterinburg, 620016, Russian Federation}
\author{A.~U.~B.~Wolter}\affiliation{Leibniz-Institut f\"ur Festk\"orper- und Werkstoffforschung (IFW) Dresden, D-01171 Dresden, Germany}
\author{S.~Wurmehl}\affiliation{Leibniz-Institut f\"ur Festk\"orper- und Werkstoffforschung (IFW) Dresden, D-01171 Dresden, Germany}\affiliation{Institut f\"ur Festk\"orperphysik, Technische Universit\"at Dresden, D-01062 Dresden, Germany}
\author{R.~Khasanov}\affiliation{Laboratory for Muon Spin Spectroscopy, Paul Scherrer Institut, CH-5232 Villigen PSI, Switzerland}
\author{I.~Eremin}\affiliation{Institut f\"ur Theoretische Physik III, Ruhr-Universit\"at Bochum, D-44801 Bochum, Germany}
\author{B.~B\"uchner}\affiliation{Leibniz-Institut f\"ur Festk\"orper- und Werkstoffforschung (IFW) Dresden, D-01171 Dresden, Germany}\affiliation{Institut f\"ur Festk\"orperphysik, Technische Universit\"at Dresden, D-01062 Dresden, Germany}

\widetext

\begin{abstract}
The superconducting properties of LaFeAsO$_{1-x}$F$_{x}$ in conditions of optimal electron-doping are investigated upon the application of external pressure up to $\sim 23$ kbar. Measurements of muon-spin spectroscopy and dc magnetometry evidence a clear mutual independence between the critical temperature $T_{c}$ and the low-temperature saturation value for the ratio $n_{s}/m^{*}$ (superfluid density over effective band mass of Cooper pairs). Remarkably, a dramatic increase of $\sim 30$ \% is reported for $n_{s}/m^{*}$ at the maximum pressure value while $T_{c}$ is substantially unaffected in the whole accessed experimental window. We argue and demonstrate that the explanation for the observed results must take the effect of non-magnetic impurities on multi-band superconductivity into account. In particular, the unique possibility to modify the ratio between intra-band and inter-bands scattering rates by acting on structural parameters while keeping the amount of chemical disorder constant is a striking result of our proposed model.
\end{abstract}


\date{\today}

\maketitle

\narrowtext

Experimental evidences together with the currently accepted theoretical models ascribe the wealth of electronic ground states exhibited by $R$FeAsO$_{1-x}$F$_{x}$ ($R$: rare-earth ion), and more generally by all iron-based pnictides, to a complex and hard-to-disentangle interplay of chemical charge doping and multi-orbital degrees of freedom,\cite{Chu08,Maz09,Mai11b,Chu12,Fer14,Yu14,Bae14,Ere14} quenched disorder\cite{Ban09,Ona09,Vor09,Efr11,Vav11,Fer12,Pra13b} and lattice strains.\cite{Gar08,Kur09,Gar11,Kan14} In this respect, an interesting alternative to modify the behaviour of pnictides in a more controlled way, i.~e., by keeping the chemical levels of charge doping and quenched disorder constant, is to apply an external pressure ($P$). $R$FeAsO$_{1-x}$F$_{x}$ compounds have been shown indeed to be a fertile ground to investigate the effects of $P$ across the whole electronic phase diagram.\cite{DeR12,Kha11,Miy13,Pra13a,Pra14,McE14} Accordingly, $P$ may be employed as a crucial tool in order to clarify important issues in these materials concerning, in particular, the exotic nature of the superconducting ground state. As a well-known result in the study of high-$T_{c}$ superconductivity (SC), a universal linear scaling is reported for the critical temperature $T_{c}$ vs. $n_{s}(0)/m^{*}$, i.~e., the ratio between the low-temperature ($T$) saturation value of the superfluid density $n_{s}$ and the effective band mass $m^{*}$ of Cooper pairs. The so-called Uemura relation $T_{c} \propto n_{s}(0)/m^{*}$ is obeyed indeed by several materials quantitatively, i.~e., with the same slope, giving strong evidence for a breakdown of the weak-coupling scenario and, in turn, for unconventional SC.\cite{Uem88,Uem89,Uem91,Uem04} The first investigations of $n_{s}$ in $R$FeAsO$_{1-x}$F$_{x}$ seemed to reconcile them well to other superconductors.\cite{Lue08,Car09} However, more extensive reports on several samples of different origin show indeed that $R$FeAsO$_{1-x}$F$_{x}$ superconductors do not obey the Uemura relation.\cite{Car13} As these results may be critically influenced by the different conditions of doping and quenched disorder in the considered samples, clarifying measurements limiting the effect of chemical perturbations are needed for the aim of more consistent results.

In this paper, we report on an investigation of the effect of $P \lesssim 23$ kbar in an optimally electron-doped sample of LaFeAsO$_{1-x}$F$_{x}$. Our results of dc magnetometry confirm the absence of a positive effect of $P$ on $T_{c}$.\cite{Miy13} At the same time, we point out by means of muon spin rotation ($\mu^{+}$SR) that SC is remarkably reinforced by $P$, as reflected in a dramatic enhancement of $n_{s}(0)/m^{*}$ by $\sim 30$ \% at the maximum $P$ value. Such a dramatic increase under $P$ while keeping $T_{c}$ constant is an unprecedented observation for any superconducting material, to the best of our knowledge. We provide evidence from density-functional theory (DFT) calculations that this result should not be associated to an induced change in the fermiology of LaFeAsO$_{1-x}$F$_{x}$ or, equivalently, to a $P$ dependence of $m^{*}$. We argue that the observed effect can only emerge in a multi-band superconductor in the presence of non-magnetic impurities such as, e.~g., As vacancies.\cite{Ham10} Strikingly, the mutual independence of $T_{c}$ and $n_{s}(0)$ can be explained within a multi-band model by assuming that $P$ modifies the ratio between intra-band and inter-bands impurity scattering rates by only acting on structural parameters while keeping the amount of chemical quenched disorder constant.

\begin{figure}[t!]
\vspace{5.7cm} \includegraphics{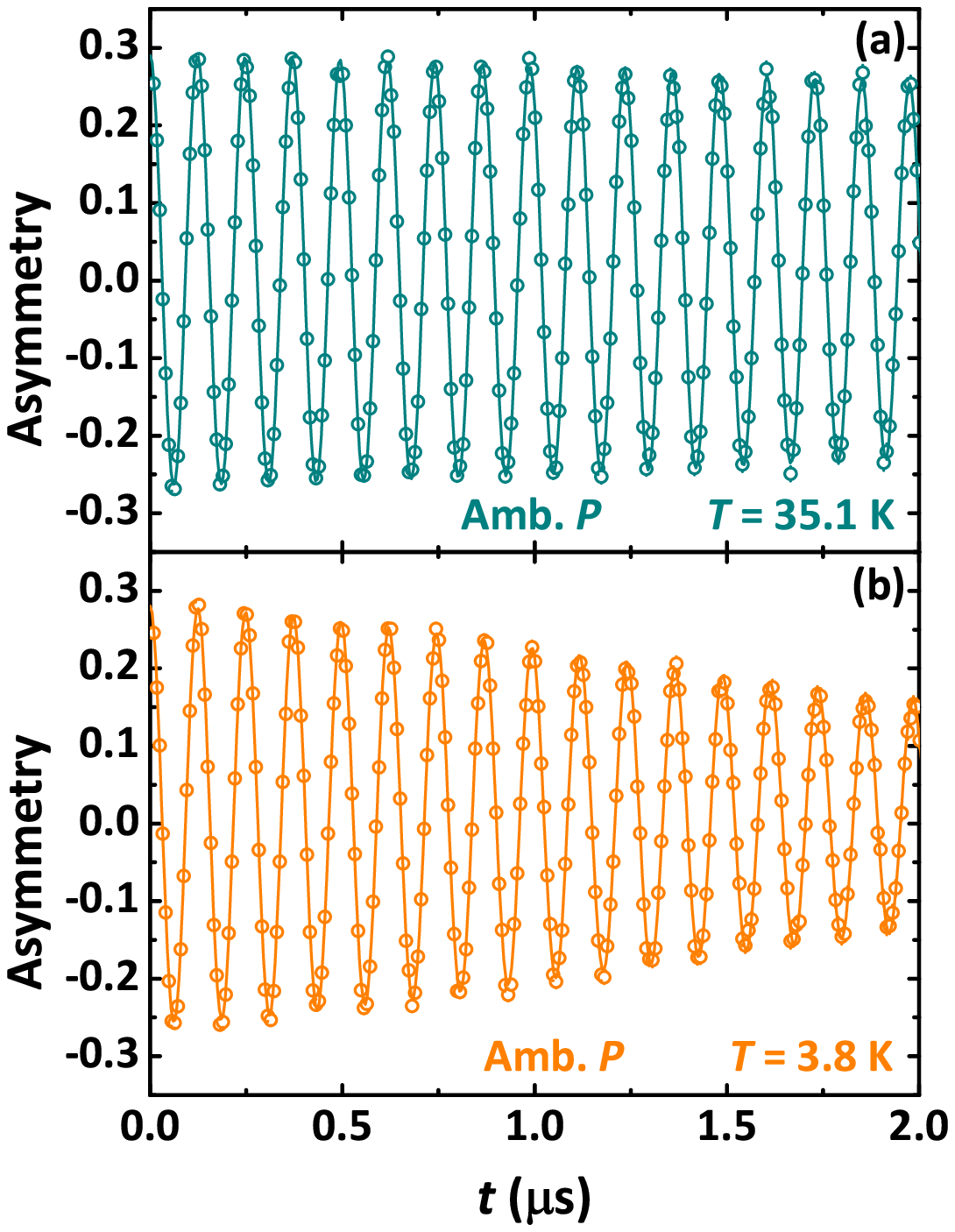} \includegraphics{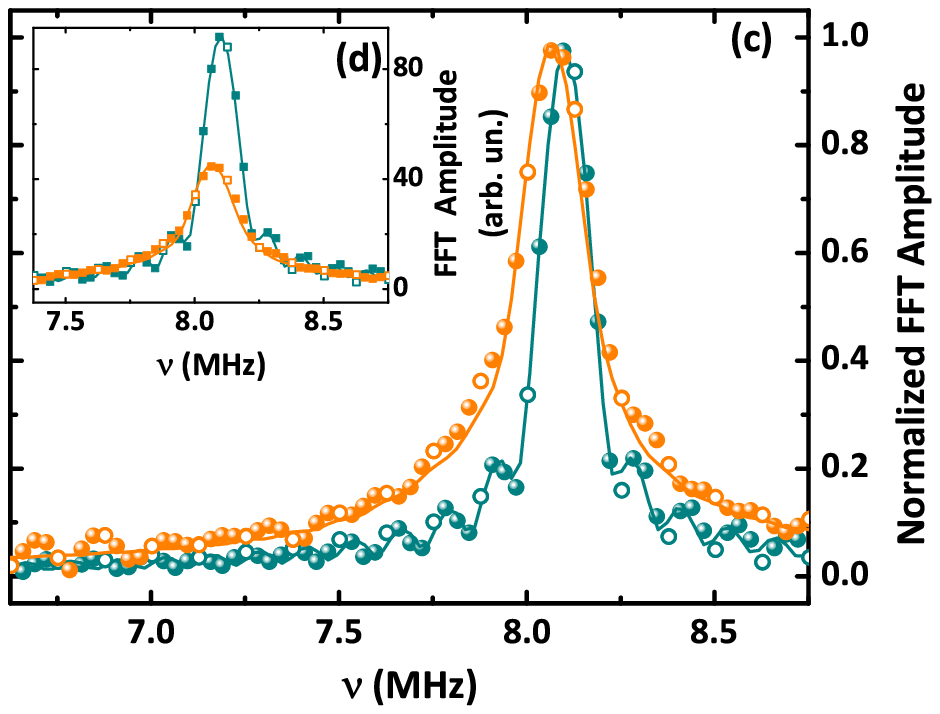}
\caption{\label{GraRawDepolarizationInCell} Open points in panels (a) and (b) are experimental TF-$\mu^{+}$SR depolarization curves at representative $T$ values well above and well below $T_{c} \simeq 24$ K, respectively (data are relative to ambient $P$ and $B_{ext} = 600$ G), while solid lines are time-domain fitting curves (see text). FFT are performed with no preliminary apodization and the normalized spectra of experimental data in panels (a) and (b) are reported in panel (c) as open symbols using the same colour code [non-normalized FFT spectra are displayed in panel (d)]. Solid symbols in panels (c) and (d) are the results of FFT after a preliminary zero-padding extending up to $4$ times the experimental window. FFT of fitting curves in panels (a) and (b) are reported as solid lines in panels (c) and (d) after zero-padding extending up to $4$ times the experimental window.}
\end{figure}
The impact of $P$ on $n_{s}$ was investigated in a polycrystalline sample of optimally-doped LaFeAsO$_{1-x}$F$_{x}$ by means of transverse-field (TF) $\mu^{+}$SR performed on the GPD spectrometer, $\mu$E$1$ beamline, at the Paul Scherrer Institute, Switzerland. Fig.~\ref{GraRawDepolarizationInCell} shows representative TF depolarization curves obtained for the sample loaded in the pressure cell at ambient $P$ and for $T$ values well above and well below $T_{c} \simeq 24$ K [panels (a) and (b), respectively]. The external transverse magnetic field $B_{ext} = 600$ G induces a coherent precession of implanted $\mu^{+}$ (the condition $B_{c1} \ll B_{ext} \ll B_{c2}$ holds, $B_{c1}$ and $B_{c2}$ being the lower and upper critical fields for the superconductor, respectively). For $T \gg T_{c}$, this precession is only weakly damped by the electronic magnetism of the pressure cell material and by the nuclear moments of both sample and pressure cell material, while a much higher damping is observed for $T \ll T_{c}$. The details of the time-domain fitting of our experimental data [see the continuous lines in Fig.~\ref{GraRawDepolarizationInCell}(a) and (b)] are described in the supplementary material. Fast Fourier transforms (FFT) were performed with no preliminary data-apodization in order to avoid artificial extra-broadening. FFT spectra of both experimental data and fitting curves are reported in Figs.~\ref{GraRawDepolarizationInCell}(c) and \ref{GraRawDepolarizationInCell}(d), enlightening for $T \ll T_{c}$ both a shift of the signal to lower frequencies and the extra-broadening discussed above. These are the well-known $\mu^{+}$SR signatures of type-II superconductivity. The former effect is due to the diamagnetism in the superconducting phase, while the latter is due to the modulation of $B_{\mu}$, namely the local magnetic field at the $\mu^{+}$ site, introduced by the vortex glass phase in the mixed Shubnikov state of the $H$--$T$ phase diagram. The detailed properties of this modulation are determined by the penetration depth $\lambda$, whose $T$ dependence can be derived accordingly. Asymmetric FFT spectra are typically expected for $T \ll T_{c}$ in type-II superconductors.\cite{Blu99,Son00} However, this feature is hindered in the current measurements by the polycrystalline nature of the sample and by the strong influence of the stray magnetic fields from the superconducting phase on the pressure cell.\cite{Mai11} Accordingly, the induced extra-broadening $\sigma_{sc}$ is of Gaussian character and one has indeed $\sigma_{sc}(T) \propto \lambda(T)^{-2} \propto n_{s}(T)/m^{*}$.\cite{Bra88,Pum90,Blu99,Son00,Mai11,Car13}
\begin{figure}[t!]
\vspace{5.8cm} \includegraphics{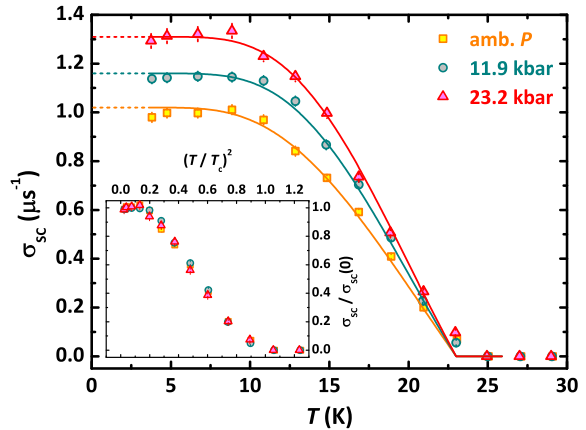}
\caption{\label{GraSuperfluidDensity} Main panel: $\sigma_{sc}$ as a function of $T$ at the different investigated $P$ values. Solid lines are best-fit curves according to a single-gap $s$-wave behaviour, whose extrapolations to the $T = 0$ limit are shown by dashed lines. Data are also reported in the inset using the same symbols as a function of $\left(T/T_{c}\right)^{2}$ and after normalization over the respective $\sigma_{sc}(0)$ values.}
\end{figure}

Results of $\sigma_{sc}$ as a function of both $T$ and $P$ for the current LaFeAsO$_{1-x}$F$_{x}$ sample are presented in Fig.~\ref{GraSuperfluidDensity}. $\sigma_{sc}(T)$ can be tracked fairly well by a single-gap $s$-wave behaviour as shown by the solid lines in Fig.~\ref{GraSuperfluidDensity}, in good agreement with previous $\mu^{+}$SR reports on similar samples.\cite{Lue08} The actual $\sigma_{sc}(T)$ dependence will not be considered any longer in this paper and its discussion will be postponed to another work. A sizeable enhancement ($\sim 30$ \%) of the saturation value $\sigma_{sc}(0)$ is clearly noticed upon increasing $P$. However it must be stressed that, as displayed in the inset of Fig.~\ref{GraSuperfluidDensity} for data normalized over the respective $\sigma_{sc}(0)$ values, no qualitative change is observed for the actual $\sigma_{sc}(T)$ shape at different $P$ values. At the same time, data in Fig.~\ref{GraSuperfluidDensity} remarkably show that $T_{c}$ is unaffected by $P$. This latter observation is independently confirmed by means of dc magnetometry, as shown in Fig.~\ref{GraMagnetization} (details in the supplementary material). Measurements of dc magnetization ($M$) were performed at $H = 10$ Oe in zero-field cooling (ZFC) conditions and results clearly enlighten that the superconducting transition of the sample is only weakly modified by $P$. At most, $T_{c}$ is even slightly shifted to lower values at the maximum $P$ value, as shown in the inset of Fig.~\ref{GraMagnetization}. The shielding fraction is also not affected at all by $P$ and it is $\sim 100$ \%, as displayed by the calibration measurement performed without the pressure cell [full symbols in Fig.~\ref{GraMagnetization}, main panel]. The trend observed for our results is in excellent agreement to what is reported in literature for a nominally identical compound.\cite{Miy13} Misleading reports on the effect of $P$ on SC in $R$FeAsO$_{1-x}$F$_{x}$ often make use of the broadened resistive SC transition to claim a dramatic increase of $T_{c}$ upon increasing $P$.\cite{Tak08a,Tak08b,Tak15} However, it should be remarked that bulk magnetic estimates of the real thermodynamic $T_{c}$ point out a much more modest effect of $P$ and, interestingly, the $T_{c}$ values upon increasing $P$ seem to be limited by the value obtained at the optimal chemical doping.\cite{Miy13}
\begin{figure}[t!]
	\vspace{6.2cm} \includegraphics{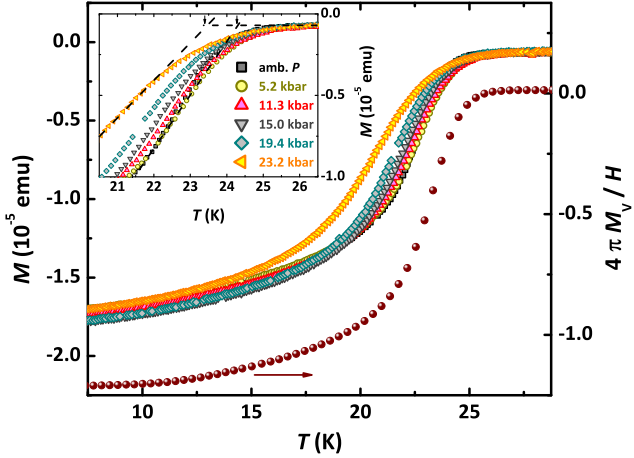}
	\caption{\label{GraMagnetization} Main panel: dc magnetization $M$ as a function of $T$ for different $P$ values (open symbols, left scale). Measurements were performed in ZFC conditions ($H = 10$ Oe) and a reference curve was taken in ZFC conditions ($H = 5$ Oe) with the sample without the pressure cell (solid symbols, right scale -- $1/4 \pi$ units). An enlargement of data in the region of the superconducting transition is displayed in the inset, where dashed lines define $T_{c}$ values, indicated by arrows.}
\end{figure}

Our experimental findings unambiguously show that $T_{c}$ and $\sigma_{sc}(0)$ are uncorrelated quantities upon increasing $P$, clearly implying a breakdown of the Uemura relation in LaFeAsO$_{1-x}$F$_{x}$. In the case of cuprates, $T_{c} \propto n_{s}(0)/m^{*}$ has been confirmed in the underdoped region of the electronic phase diagram.\cite{Uem88,Uem89,Uem91,Uem04} However, clear deviations have been observed in the past, the probably most remarkable one being reported for Y$_{1-x}$Ca$_{x}$Ba$_{2}$Cu$_{3}$O$_{7-\delta}$ as a function of chemical substitutions. Here, a clear saturation effect for $T_{c}$ with $n_{s}$ strongly increasing was observed and explained in terms of different contributions to $n_{s}$ arising from carriers on CuO$_{2}$ planes and CuO chains.\cite{Tal95} A similar effect was measured in YBa$_{2}$Cu$_{3}$O$_{7-\delta}$ upon increasing $P$.\cite{Mai11} However, the relative enhancement of $n_{s}$ under $P$ is $\sim 3$ times smaller than what is observed in the current case for LaFeAsO$_{1-x}$F$_{x}$ and, at the same time, the explanation can still rely on charge contributions from planes and chains to $n_{s}$, an argument which is not suited at all to LaFeAsO$_{1-x}$F$_{x}$. These issues denote deep intrinsic distinctions indicative of a different underlying physics and, accordingly, novel scenarios should be considered to correctly understand our observations. 

\begin{figure}[t!]
	\vspace{2.9cm} \includegraphics{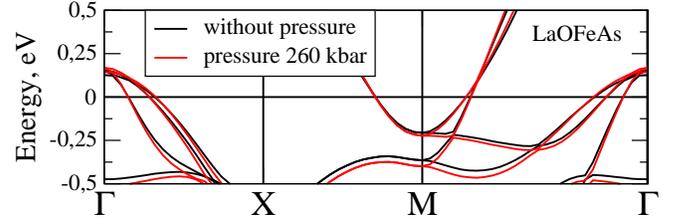}
	\caption{\label{GraBands} Band structure of LaFeAsO at ambient $P$ (black curves) and for $P = 260$ kbar (red curves).}
\end{figure}
As discussed above, the measured extra-broadening $\sigma_{sc}$ of the $\mu^{+}$SR spectrum mainly depends on two contributions, namely $n_{s}$ and $m^{*}$. However, a strong $P$ dependence of $m^{*}$ can be ruled out by DFT calculations of the electronic band structure. In particular, as shown in Fig.~\ref{GraBands}, electronic bands (and their inverse second derivatives) computed in the local-density approximation (LDA) within the linearised muffin-tin orbitals method using default settings are only weakly affected by $P$ even up to $260$ kbar, namely much beyond the experimentally accessible range. This is in agreement with previous reports for lower $P$ values.\cite{DeR12} Including low-energy electronic correlations beyond LDA will certainly introduce a moderate renormalization of $m^{*}$, yet these correlations will typically also affect $T_{c}$ at the same time due to the modification of the low-energy part of the bosonic spectrum. Accordingly, to understand the experimental results on a quantitative level, we focus directly on the intrinsic density of supercarriers $n_{s}$. We start from a model for the $s^{\pm}$ superconductor which consists of superconducting gaps of similar magnitudes but of opposite signs formed on hole ($h$) and electron ($e$) bands.\cite{Chu08,Maz09,Umm09,Umm11} The electron pockets are located around the $X = (\pi , 0)$ points of the Brillouin zone, which have lower symmetry than the $\Gamma = (0 , 0)$ or the $M = (\pi , \pi)$ points, where hole pockets are located. Accordingly, we allow for a $\cos 2 \theta$ modulation of the gap on the electron pocket\cite{Yu14}
\begin{equation}\label{trueformofDeltae}
\Delta_{e}(\theta) = \Delta_{e} \pm \Delta_{\bar{e}} \cos 2\theta \qquad (\textrm{where} \; \Delta_{e} = -\Delta_{h}).
\end{equation}
A modification of the mediating boson responsible for SC under $P$, i.~e., $P$-dependent intra-band ($\lambda_{ee}$ and $\lambda_{hh}$) and inter-bands ($\lambda_{he}$) couplings, may well-affect $n_{s}(0)$. However, the modification of $\lambda$ couplings alone is not enough to justify our results, as this would simultaneously modify $T_{c}$ as well.

Following the discussion above, we argue that a mutual independence of $n_{s}(0)$ and $T_{c}$ with increasing $P$ may only be obtained by taking the effect of scattering from non-magnetic impurities into account. Due to the internal sign change of the gap between the $h$ and $e$ pockets, the effect of such impurities can be separated into two contributions. The intra-band scattering rate $\Gamma_{0}$ within $h$ and/or $e$ pockets does not affect the superconducting gap and $T_{c}$ for $s$-wave symmetry.\cite{Abr59a,Abr59b,And59,Abr94} On the other hand, the inter-bands scattering rate $\Gamma_{\pi}$ between $h$ and $e$ pockets leads to strong pair-breaking effects, similarly to the case of magnetic scattering.\cite{Gol97,Vor09,Vav11} In particular, while a change of $\Gamma_{0}$ only affects the superfluid density and keeps $T_{c}$ constant, $\Gamma_{\pi}$ affects both quantities in a similar fashion.\cite{Vor09} Non-magnetic impurities change the balance between the angle-independent and dependent parts of the gap on the $e$ pockets as well. In the presence of impurities, one identifies the new Matsubara frequencies $i\tilde{\omega}_{m}^{h} = i\omega_{m}\eta_{m}^{h}$, and $i\tilde{\omega}_{m}^{e} = i\omega_{m}\eta_{m}^{e}$ for $h$ and $e$ pockets, respectively, as well as the superconducting gaps $\tilde{\Delta}_{m}^{h} = \bar{\Delta}_{m}^{h}\eta_{m}^{h}$ and $\tilde{\Delta}_{m}^{e}(\theta) = \bar{\Delta}_{m}^{e}(\theta)\eta_{m}^{e}$. All these quantities are determined self-consistently for a given temperature and impurity scattering as described in the supplementary material. Finally, the expression for $n_{s}(T)$, i.~e., the zero-frequency value of the current-current correlation function, can be written as
\begin{eqnarray}\label{superfluidfertig}
\frac{n_{s}(T)}{n_{s}(0)} & = &\frac{\pi T}{2}\sum_{\omega_{m}} \left[\frac{|\bar{\Delta}_{m}^{h}|^{2}}{\eta_{m}^{h}(\omega_{m}^{2}+|\bar{\Delta}_{m}^{h}|^{2})^{3/2}} \right. \nonumber\\ & & \left. +\left\langle\frac{|\bar{\Delta}_{m}^{e}|^{2}}{\eta_{m}^{e}(\omega_{m}^{2}+|\bar{\Delta}_{m}^{e}(\theta)|^{2})^{3/2}}\right\rangle_\theta \right].%
\end{eqnarray}
\begin{figure}[t!]
	\vspace{3.35cm} \includegraphics{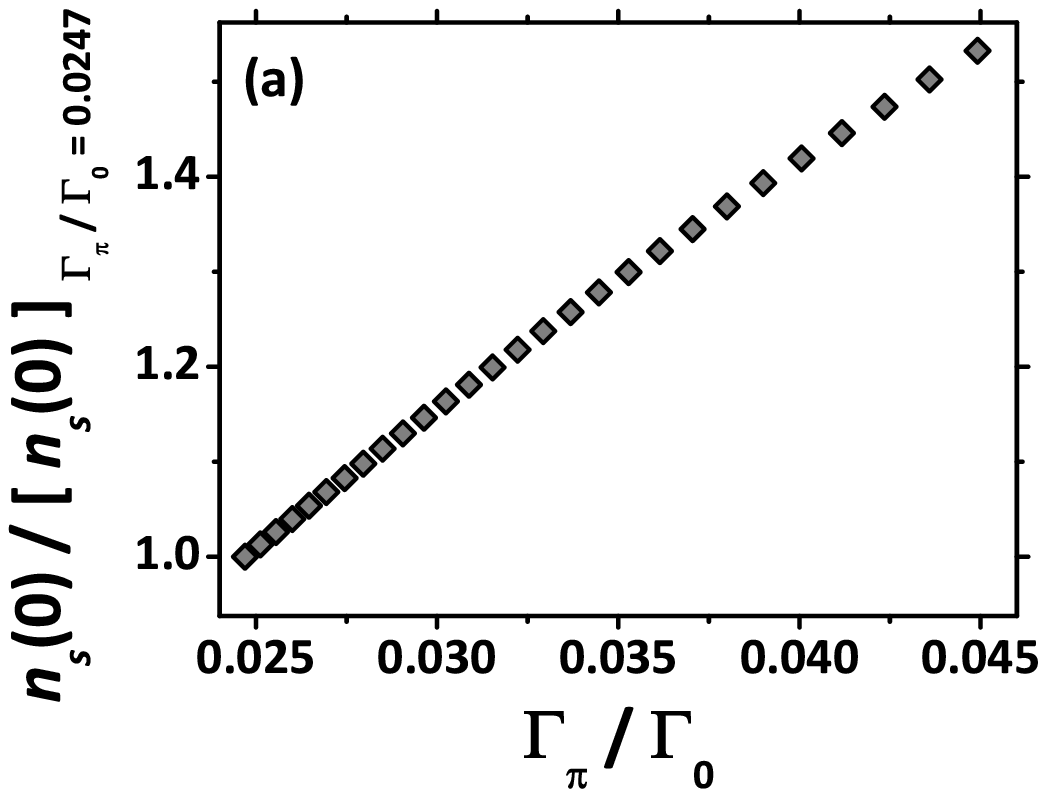} \includegraphics{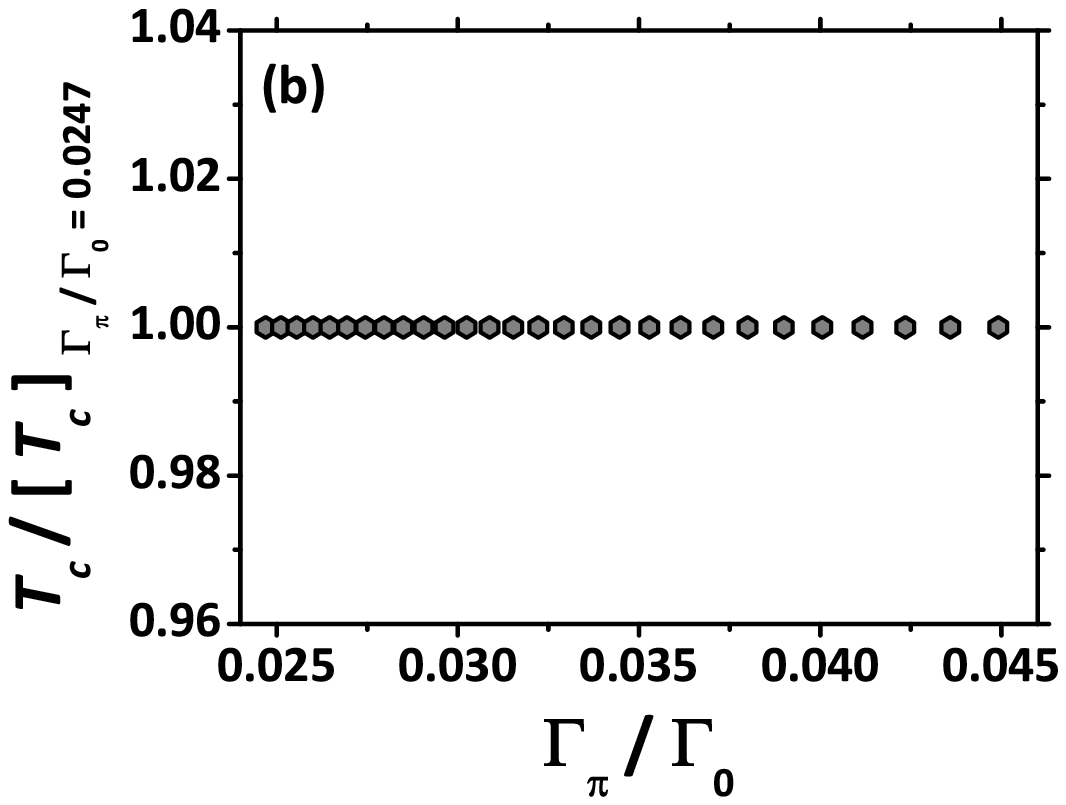}
	\caption{\label{fig_theory} Calculated behaviour of $n_{s}(0)$ and $T_{c}$ [panels (a) and (b), respectively] as a function of the $\Gamma_{\pi}/\Gamma_{0}$ ratio. Both the quantities are normalized to their values at $\Gamma_{\pi}/\Gamma_{0} = 0.0247$. We further employ $m_{h} = 2.0 m$ and $m_{e} = 1.5 m$ with energy dispersions $\xi_{h} = - k^{2}/(2 m_{h}) + \mu$, $\xi_{e} = k^{2}/(2 m_{e}) - \mu - \mu_{0}$. Here, $\mu = 3$ and $\mu_{0} = 0.2$.}
\end{figure}

To model our experimental results, we then assume that the multi-band nature of SC in pnictides allows $P$ to modify the $\Gamma_{\pi}/\Gamma_{0}$ ratio while still keeping the chemical density of non-magnetic impurities constant. It should be pointed out that significant changes of $\Gamma_{\pi}$ with increasing $P$ would affect the actual $T$ dependence of $n_{s}$,\cite{Vor09} opposite to the experimental findings reported in Fig.~\ref{GraSuperfluidDensity}. Therefore, the limit $\Gamma_{\pi} \ll \Gamma_0$ is considered. The striking result of our calculations is that, under these circumstances, a relatively small variation of the $\Gamma_{\pi}/\Gamma_0$ ratio even in the presence of a constant amount of impurities is indeed able to reproduce well the experimental data, as shown in Fig.~\ref{fig_theory} where a clear decoupling of $n_{s}$ and $T_{c}$ is reported. It must be noticed that such a behaviour would be characteristic for any multi-band superconductor independently on the phase structure of the superconducting gap on the difference Fermi surfaces, namely for both $s^{\pm}$ and $s^{++}$. It is also worth stressing that assuming a modification in $\Gamma_{\pi}/\Gamma_0$ implicitly involves a $P$ dependence of the orbital composition of the $t_{2g}$ states at the Fermi energy and, accordingly, an important orbital character of superconductivity. While these arguments have already been discussed on a more quantitative basis for other pnictide superconductors at ambient $P$,\cite{Yu14} nothing is directly known for $R$FeAsO$_{1-x}$F$_{x}$ due to the current lack of ARPES data. As such, our results give an interesting hint towards an universal behaviour shared by different families of pnictides.

Summarizing, we have reported on the superconducting properties of optimally electron-doped LaFeAsO$_{1-x}$F$_{x}$ under $P$. A breakdown of the Uemura relation is unambiguously evidenced, with $T_{c}$ unaffected by $P$ and an increase of $n_{s}(0)/m^{*}$ by $\sim 30$ \% at $P \simeq 23$ kbar. We propose an explanation based on the unique possibility to modify the ratio between intra-band and inter-bands scattering rates in multi-band superconductors by only acting on structural parameters while keeping the amount of chemical disorder constant. This scenario can be accounted for by assuming an important orbital character of superconductivity in LaFeAsO$_{1-x}$F$_{x}$.

We acknowledge useful and stimulating discussions with A. V. Chubukov and G. A. Ummarino. G. P. acknowledges support by the Humboldt Research Fellowship for Postdoctoral Researchers. The work of T. H., F. A., S. W., I. E. and B. B. is supported by the DFG under the Priority Program SPP 1458. I. E, T. H. and F. A. acknowledge support by the German Academic Exchange Service (DAAD PPP USA No. 57051534). S. S. acknowledges partial support of PRIN2012 Project No. 2012X3YFZ2. I. N. acknowledges support by the RFBR grant No. 14-02-00065. S. W. acknowledges funding by DFG in the Emmy Noether project WU595/3-2. I. E. thanks the Platform for Superconductivity and Magnetism Dresden for its kind hospitality. We thank R. Wachtel, S. M\"uller-Litvanyi and G. Kreutzer for technical support.



\newpage

\widetext

\section*{SUPPLEMENTARY MATERIAL}

\section{Muon-Spin Spectroscopy ($\bm{\mu^{+}}$SR)}

Measurements of zero- and transverse-field (ZF and TF, respectively) $\mu^{+}$SR were performed at the Paul Scherrer Institute on the GPS and GPD spectrometers ($\pi$M$3$ and $\mu$E$1$ beamlines, respectively) for $1.6$ K $< T < 200$ K. In a $\mu^{+}$SR experiment, a beam of spin-polarized positive muons $\mu^{+}$ is implanted in well-defined crystallographic sites of the investigated material.\cite{Blu99,Yao11} The time ($t$) dependence of the spin (de)polarization of $\mu^{+}$, $P_{T}(t) = A_{T}(t)/A_{0}$, can be monitored as a function of temperature ($T$) and, accordingly, detailed information can be achieved about the local magnetic features characteristic of the material. Here, $A_{T}(t)$ is the so-called asymmetry function (see below) while $A_{0}$ is an experimental instrument-dependent parameter. In the case of superconductors, TF-$\mu^{+}$SR is one of the most suited techniques in order to directly quantify the penetration length $\lambda$ and, for single-crystalline and high-purity samples, to investigate the symmetry of the superconducting gap in detail.\cite{Blu99,Yao11,Har87,Bra88,Pum90,Son94,Son00,Car13} Differently from, e.~g., inductive techniques, TF-$\mu^{+}$SR has then the great advantage of allowing one to properly achieve the absolute value of $\lambda$ and not only its variations.

Data obtained in the low-background spectrometer GPS are used as reference for the more subtle analysis performed on data under $P$ (spectrometer GPD). Here, the sample is inserted into a double-wall pressure cell (PC) made up of MP$35$N alloy. $P$ is triggered by pistons of MP$35$N alloy, while the transmitting medium is Daphne oil $7373$, ensuring almost-optimal conditions of hydrostaticity up to $P \sim 2$ GPa.\cite{Yok07,Kha11} The actual value of $P$ is measured at low-$T$ by checking the superconducting transition temperature of a small indium manometer inside the cell by means of ac susceptibility. Only a fraction $\sim 35$ \% of the incoming $\mu^{+}$ is actually implanted into the sample, all the others being stopped by the PC. For this reason, it is crucial to characterize the magnetic behaviour of the cell in a separate set of TF-$\mu^{+}$SR measurements in the same experimental conditions. In the case of superconducting materials for $T < T_{c}$, a stray magnetic field is induced outside the sample when an external field is applied. Accordingly, also $\mu^{+}$ implanted into the PC and close enough to the sample will probe a distribution $P(B_{\mu})$ of local magnetic field resulting in a corresponding damping of the relative TF-$\mu^{+}$SR signal. Accordingly, the typical fitting function employed in order to analyse data is written as\cite{Mai11}
\begin{eqnarray}
\frac{A_{T}^{\textrm{TF}}(t)}{A_{0}} & = & a_{s} \exp\left[-\frac{\left(\sigma_{n}^{2}+\sigma_{sc}^{2}\right) t^{2}}{2}\right] \cos\left(\gamma B_{\mu} t + \phi\right) \nonumber \\ & & + \left(1-a_{s}\right) \exp\left[-\left(\frac{\sigma_{PC}^{2} t^{2}}{2}+\lambda_{PC}t\right)\right] \times \int P(B_{\mu}^{\prime}) \cos\left(\gamma B_{\mu}^{\prime} t + \phi\right) d B_{\mu}^{\prime}.
\end{eqnarray}
The parameter $a_{s}$ is introduced in order to quantify the fraction of $\mu^{+}$ implanted into the sample [specular to $\left(1-a_{s}\right)$ which accounts for $\mu^{+}$ implanted into the PC], while $B_{\mu}$ quantifies the local magnetic field probed by $\mu^{+}$ (the resulting oscillation is phase-shifted by the factor $\phi$ while $\gamma = 2 \pi \times  135.54$ MHz/T represents the gyromagnetic ratio of $\mu^{+}$). The $T$-independent Gaussian relaxation rate $\sigma_{n}$ accounts for nuclear magnetism in the sample, while the $T$-dependent Gaussian contribution $\sigma_{sc}$ is the main outcome of the experiment, namely the extra-damping induced by the vortices. The relaxation rates $\sigma_{PC}$ and $\lambda_{PC}$ describe the intrinsic magnetic properties of the PC, where the distribution $P(B_{\mu})$ is induced by the stray magnetic field from the sample.

The actual value of the external TF was chosen after performing a $H$-scan at $T = 1.6$ K aiming at maximizing the value of $\sigma_{SC}$. The overall $\sigma_{SC}$ vs. $H$ trend is in excellent quantitative agreement with previously reported data.\cite{Lue08} Accordingly, the value $H = 600$ Oe was selected for the $T$ scans, both in GPS and in GPD. 

\section{dc Magnetization}

The diamagnetic response of the sample was carefully studied by means of dc magnetometry in zero-field cooling (ZFC) conditions and upon the application of $P$. In particular, as discussed in the paper, the following information was extracted about superconductivity, namely $T_{c}$, transition width and superconducting shielding fraction. Measurements were performed in a commercial superconducting quantum interference device (SQUID) magnetometer MPMS-XL$5$ (Quantum Design). A homemade PC was employed, whose design is similar to that of a diamond anvil cell. Two opposing cone-shaped ceramic anvils compress a gasket with a small hole that serves as a sample chamber. The applied uniaxial pressure onto the gasket is transformed into hydrostatic pressure in the sample chamber by using Daphne oil $7373$ as transmitting medium. A single powder-grain is used as a sample. The actual $P$ value is measured by monitoring the superconducting transition temperature of a small Pb manometer inside the sample chamber. All the mechanical parts are made of weakly-magnetic materials allowing a dramatic reduction of the background signal even in the (current) case of samples with tiny dimensions.

\section{Non-magnetic impurity scattering in a $s^{\pm}$ superconductor}

Here, we present the derivation of the superconducting gap equations for the $s^{\pm}$ superconductor in presence of non-magnetic impurities, following Ref.~\onlinecite{Vor09}. Here, the superconducting gaps are formed on the electron and on the hole bands of similar magnitudes but of opposite signs. In addition, the electron pockets are located around the $(\pi,0)$ points of the BZ, which have lower symmetry than the $\Gamma$-point or the  $M = (\pi,\pi)$ points of the BZ, where hole pockets are located. As a result, one should allow the $\cos 2 \theta$ modulation of the gap on the electron pocket. Due to internal sign change of the gap between the electron and the hole pockets the effect of the non-magnetic impurity can be separated into the scattering within either hole and/or electron pockets (denoted as $\Gamma_{0}$), which does not influence the superconducting gap and $T_{c}$ and the scattering between electron and hole pockets (here $\Gamma_{\pi}$), which behaves as magnetic impurity in this case with strong pair-breaking effects.

To find the gap equations we write the first two \textit{Gorkov equations} for the hole and the electron pockets
\begin{alignat}{2}
	&[i\omega_m+\Gamma_0\bar{G}_h+\Gamma_\pi\bar{G}_e-\xi_h]G_h&+[\Delta_h+\Gamma_0 \bar{F}_h+\Gamma_\pi\bar{F}_e]F_h^\dagger&=1,\nonumber\\
	&[i\omega_m+\Gamma_0\bar{G}_e+\Gamma_\pi\bar{G}_h-\xi_e]G_e&+[\Delta_e(\theta)+\Gamma_0\bar{F}_e+\Gamma_\pi\bar{F}_h]F_e^\dagger&=1
\end{alignat}
where
\begin{alignat*}{2}
	G_h&= G_h(\tilde{\Delta}_m^h,\tilde{\omega}_m^h)&,\quad F_h&= F_h(\tilde{\Delta}_m^h,\tilde{\omega}_m^h),\nonumber\\
	G_e&= G_e(\tilde{\Delta}_m^e,\tilde{\omega}_m^e)&,\quad F_e&= F_e(\tilde{\Delta}_m^e,\tilde{\omega}_m^e)
\end{alignat*}
are the normal and the anomalous Green's function for the hole and the electron pockets, and $\bar{G}_h$ and $\bar{G}_e$ satisfy the equations
\begin{align*}
	\bar{G}_h&=\frac{1}{\pi}\int d\xi_h G_h,\nonumber\\
	\bar{G}_e&=\frac{1}{\pi}\int d\xi_e G_e.
\end{align*}
In this model, the ansatz for the $s^{\pm}$-wave gap is
\begin{align}
	\Delta_e(\theta)&=\Delta_e\pm\Delta_{\bar{e}}\cos(2\theta)
	\label{trueformofDeltae}
\end{align}
and $\Delta_h=-\Delta_e$. In the presence of impurities one defines the new Matsubara frequencies,
\begin{alignat}{2}
	i\tilde{\omega}_m^h&=i\omega_m\eta_m^h&=i\omega_m+\Gamma_0\bar{G}_h+\Gamma_\pi\bar{G}_e,\nonumber\\
	i\tilde{\omega}_m^e&=i\omega_m\eta_m^e&=i\omega_m+\Gamma_0\bar{G}_e+\Gamma_\pi\bar{G}_h,\label{eq:renorm_freqs}
\end{alignat}
as well as the superconducting gaps
\begin{alignat}{2}
	\tilde{\Delta}_m^h&=\bar{\Delta}_m^h\eta_m^h&=&\Delta_h+\Gamma_0\bar{F}_h+\Gamma_\pi\bar{F}_e,\nonumber\\
	\tilde{\Delta}_m^e(\theta)&=\bar{\Delta}_m^e(\theta)\eta_m^e&=&\Delta_e(\theta)+\Gamma_0\bar{F}_e+\Gamma_\pi\bar{F}_h. \label{eq:renorm_gaps}
\end{alignat}
In Eq.~\eqref{eq:renorm_gaps}, expressions for $\bar{\Delta}_m^h$ and $\bar{\Delta}_m^e$ can be written with the help of $\eta_m^h$ and $\eta_m^e$ that are the renormalization factors for hole and electron pockets and introduce $\bar{\Delta}_h=-\bar{\Delta}_m^e$ and $\bar{\Delta}_m^e(\theta)=\bar{\Delta}_m^e\pm\bar{\Delta}_m^{\bar{e}}\cos(2\theta)$. Then, the first two \textit{Gorkov equations} can be written as
\begin{alignat*}{2}
	&(i\tilde{\omega}_m^h-\xi_h)G_h +\tilde{\Delta}_h  F_h^\dagger&=1,\nonumber\\
	&(i\tilde{\omega}_m^e-\xi_e)G_e +\tilde{\Delta}_e(\theta) F_e^\dagger&=1
\end{alignat*}
where	
\begin{align*}
	G_{e}=\frac{i\tilde{\omega}^e_m+\xi_e}{(\tilde{\omega}^e_m)^2+\xi_e^2+|\tilde{\Delta}^e_m(\theta)|^2},\:\:\:
	G_{h}=\frac{i\tilde{\omega}^h_m+\xi_h}{(\tilde{\omega}^h_m)^2+\xi_h^2+|\tilde{\Delta}^h_m|^2}
\end{align*}
are the normal Green's function for the electron and hole bands. Substituting $G_h$ and $G_e$ in Eq.~\eqref{eq:renorm_freqs} we find
\begin{align}
	\eta_m^h&=1+\Gamma_0\frac{1}{\sqrt{{({\omega}_m^h)}^2+|\bar{\Delta}_m^h|^2}} %
	+\Gamma_\pi\left\langle\frac{1}{\sqrt{{({\omega}_m^e)}^2+|\bar{\Delta}_m^e(\theta)|^2}}\right\rangle_\theta
	\label{etaheq}
\end{align}
and $\langle\cdot\rangle_{\theta}=\tfrac{1}{2\pi}\int_0^{2\pi}d\theta$ is the average with respect to the angle $\theta$. In a similar way one obtains $\eta_m^e$
\begin{align}
	\eta_m^e&=1+\Gamma_0\left\langle\frac{1}{\sqrt{{({\omega}_m^e)}^2+|\bar{\Delta}_m^e(\theta)|^2}}\right\rangle_\theta %
	+\Gamma_\pi\frac{1}{\sqrt{{({\omega}_m^h)}^2+|\bar{\Delta}_m^h|^2}}.
\end{align}
Next, one derives in a similar fashion a set of self-consistent gap equations
\begin{align}
	\begin{split}
		\bar{\Delta}_m^h&=\Delta_h-\Gamma_\pi(\bar{\Delta}_m^h-\bar{\Delta}_m^e) %
		\left\langle\frac{1}{\sqrt{\omega_m^2+|\bar{\Delta}_m^e(\theta)|^2}}\right\rangle_\theta,\\
		\bar{\Delta}_m^e&=\Delta_e%
		-\Gamma_\pi(\bar{\Delta}_m^e-\bar{\Delta}_m^h) %
		\frac{1}{\sqrt{\omega_m^2+|\bar{\Delta}_m^h|^2}},\\
		\bar{\Delta}_m^{\bar{e}}&=\Delta_{\bar{e}}%
		-\Gamma_\pi\bar{\Delta}_m^{\bar{e}} %
		\frac{1}{\sqrt{\omega_m^2+|\bar{\Delta}_m^h|^2}} %
		-\Gamma_0\bar{\Delta}_m^{\bar{e}}%
		\left\langle\frac{1}{\sqrt{\omega_m^2+|\bar{\Delta}_m^e(\theta)|^2}}\right\rangle_\theta.
	\end{split}
	\label{deltabarfertig}
\end{align}

Note that the equations for the bare superconducting gaps in the clean case without impurities ${\Delta}_h$, ${\Delta}_e$ and ${\Delta}_{\bar{e}}$ are given by
\begin{align}
	\begin{split}
		\Delta_h=&%
		-\pi T\sum_{\omega_m}^\Lambda\Biggl[\sum_{k^\prime,\theta^\prime}^{|\xi_h(k^\prime)|<\omega_D} %
		\frac{2U_{hh}\bar{\Delta}_m^h}{\omega_m^2+\xi_h(k^\prime)^2+|\bar{\Delta}_m^h|^2}\\
		&\qquad\qquad\qquad\qquad\qquad\:+\sum_{k^\prime,\theta^\prime}^{|\xi_e(k^\prime)|<\omega_D} %
		\frac{2U_{he}\bar{\Delta}_m^e+4U_{he}\alpha_{he}\bar{\Delta}_m^{\bar{e}}\cos^2{2{\theta^\prime}}} %
		{\omega_m^2+\xi_e(k^\prime)^2+|\bar{\Delta}_m^e({\theta^\prime})|^2}\Biggr],\\
		\Delta_e=&%
		-\pi T\sum_{\omega_m}^\Lambda\Biggl[\sum_{k^\prime,\theta^\prime}^{|\xi_h(k^\prime)|<\omega_D} %
		\frac{2U_{he}\bar{\Delta}_m^h}{\omega_m^2+\xi_h(k^\prime)^2+|\bar{\Delta}_m^h|^2}\\
		&\qquad\qquad\qquad\qquad\qquad\:+\sum_{k^\prime,\theta^\prime}^{|\xi_e(k^\prime)|<\omega_D} %
		\frac{2U_{ee}\bar{\Delta}_m^e+4U_{ee}\alpha_{ee}\bar{\Delta}_m^{\bar{e}}\cos^2{2{\theta^\prime}}} %
		{\omega_m^2+\xi_e(k^\prime)^2+|\bar{\Delta}_m^e({\theta^\prime})|^2}\Biggr],\\
		\Delta_{\bar{e}}=&%
		-\pi T\sum_{\omega_m}^\Lambda\Biggl[\sum_{k^\prime,\theta^\prime}^{|\xi_h(k^\prime)|<\omega_D} %
		\frac{4U_{he}\alpha_{he}\bar{\Delta}_m^h}{\omega_m^2+\xi_h(k^\prime)^2+|\bar{\Delta}_m^h|^2}\\
		&\qquad\qquad\qquad\qquad\qquad\:+\sum_{k^\prime,\theta^\prime}^{|\xi_e(k^\prime)|<\omega_D} %
		\frac{4U_{ee}\alpha_{ee}\bar{\Delta}_m^e+8U_{ee}\beta_{ee}\bar{\Delta}_m^{\bar{e}}\cos^2{2{\theta^\prime}}} %
		{\omega_m^2+\xi_e(k^\prime)^2+|\bar{\Delta}_m^e({{\theta^\prime}})|^2}\Biggr].\\
	\end{split}
	\label{deltasfertig}
\end{align}
These equations hold for two equivalent hole pockets at the Gamma point at $k = (0,0)$ and for two circular symmetry related electron pockets at the X and Y point at $q_1=(\pi,0)$ and $q_2=(0,\pi)$. The intra- and inter-pocket interactions assumed here have the usual form\cite{Mai11b}
\begin{align}
	\begin{split}
		U_{hh_{1,2}}(\phi,\phi^\prime)&=U_{hh},\\
		U_{he_{1,2}}(\phi,{\theta^\prime})&=U_{he}(1\pm 2\alpha_{he}\cos{2{\theta^\prime}}),\\
		U_{e_ie_i}(\theta,{\theta^\prime})&= %
		U_{ee}(1\pm 2\alpha_{ee}(\cos{2\theta}+\cos{2{\theta^\prime}}) %
		+4\beta_{ee}\cos{2\theta}\cos{2{\theta^\prime}}),\\
		U_{e_1e_2, e_2e_1}(\theta,{\theta^\prime})&= %
		U_{ee}(1\pm 2\alpha_{ee}(\cos{2\theta}-\cos{2{\theta^\prime}}) %
		-4\beta_{ee}\cos{2\theta}\cos{2{\theta^\prime}})\\
	\end{split}
\end{align}
and the $s^{\pm}$ solution is guaranteed for $U^2_{he}>U_{hh} U_{ee}$. We further assume that the angular dependence of the gap on the electron pockets is not strong enough to produce accidental nodes as, experimentally, there is no evidence for them in LaFeAsO$_{1-x}$F$_{x}$. The London penetration depth $\lambda(T)$ scales as $1/\sqrt{\rho_s (T)}$, where $\rho_s (T)$ is the superfluid density. The latter is, up to a factor, the zero frequency value of the current-current correlation function and can be written in the following form for the $s^{\pm}$ superconductor
\begin{align}
	\frac{\rho_s(T)}{\rho_{s0}}=\frac{\pi T}{2}\sum_{\omega_m} %
	\frac{|\bar{\Delta}_m^h|^2}{\eta_m^h(\omega_m^2+|\bar{\Delta}_m^h|^2)^{3/2}} %
	+\left\langle\frac{|\bar{\Delta}_m^e|^2}{\eta_m^e(\omega_m^2+|\bar{\Delta}_m^e(\theta)|^2)^{3/2}}\right\rangle_\theta.%
	\label{superfluidfertig}
\end{align}

\end{document}